\documentclass[a4paper,11pt]{article}
\pdfoutput=1

\usepackage{jcappub}
\usepackage[utf8]{inputenc}
\usepackage{amsmath,mathtools}
\usepackage{tensor}
\usepackage{amssymb}
\usepackage{graphicx,epsfig,amssymb}
\usepackage{physics}
\usepackage{lipsum}  
\usepackage{bigints}
\usepackage{verbatim}
\usepackage{hyperref}
\usepackage{color}
\usepackage{multirow}
\usepackage{float}
\usepackage{array}
\usepackage{orcidlink}
\usepackage{booktabs}

\usepackage{booktabs}
\usepackage{array}
\usepackage{ragged2e}
\usepackage{makecell}

\newcolumntype{P}[1]{>{\centering\arraybackslash}p{#1}}

\begin{document}

\title{Early Dark Energy Crowds Out Interacting Dark Energy}

\author[a]{William Giar\`{e}}
\author[a]{Jeremy Sakstein}
\affiliation[a]{Department of Physics and Astronomy, University of Hawai‘i, Honolulu, HI 96822, USA}
\emailAdd{giare@hawaii.edu}
\emailAdd{sakstein@hawaii.edu}

\abstract{
We investigate whether interacting dark energy can provide additional relief to the Hubble tension once early dark energy (EDE) is present. Considering a cosmology containing EDE, an independent late-time dark energy fluid with an equation of state fixed at $w\simeq-1$ interacting with cold dark matter, and a cosmological constant, we find that the data strongly drive the model toward the non-interacting limit. The best fit effectively reduces to the standard EDE cosmology, where a non-zero EDE component provides most of the required increase in $H_0$, leaving little statistical gain for the interaction. Although energy transfer from dark matter to dark energy can raise $H_0$ further, this limited gain is rapidly outweighed by the deterioration of the fit to low-redshift observations. In particular, since the CMB tightly constrains the early-time physical matter density, any additional energy transfer requires a lower present-day matter abundance. Such low values are strongly disfavored by Type Ia supernovae. We show that this limitation becomes particularly severe once EDE has already raised $H_0$ and that the same mechanism extends to a broad class of post-recombination interactions. Allowing a dynamical dark energy equation of state of the Chevallier-Polarski-Linder form does not alter our conclusions.
}
\maketitle

\section{Introduction}
\label{sec:introduction}

The Hubble tension~\cite{Verde:2019ivm,DiValentino:2020zio,DiValentino:2021izs,Perivolaropoulos:2021jda,Schoneberg:2021qvd,Shah:2021onj,Abdalla:2022yfr,DiValentino:2022fjm,Kamionkowski:2022pkx,Giare:2023xoc,Hu:2023jqc,Verde:2023lmm,CosmoVerseNetwork:2025alb,Schoneberg:2026eys,Schoneberg:2026vaf} has motivated broad interest in physics beyond the standard $\Lambda$CDM model. Most attempts to accommodate a larger $H_0$ fall into two main classes: early-time solutions that typically aim to reduce the sound horizon, requiring a larger present-day expansion rate to preserve the CMB acoustic scale; and late-time mechanisms that instead modify the distance to the last-scattering surface.

Early-time solutions remain a viable route to substantially reducing the tension~\cite{CosmoVerseNetwork:2025alb,Schoneberg:2026eys,Schoneberg:2026vaf}.\footnote{Independent constraints from the ages of the oldest Galactic stars have been argued to challenge solutions relying solely on early-time modifications~\cite{Banik:2026dyx}, see also Ref.~\cite{Vagnozzi:2023nrq}.} Recent comparative analyses place early dark energy~\cite{Wetterich:2004pv,Doran:2006kp,Kamionkowski:2022pkx,Poulin:2025nfb} (EDE) among the best-performing phenomenological mechanisms in this class~\cite{Schoneberg:2026eys,Schoneberg:2026vaf}.\footnote{For a comprehensive overview of EDE models, including recent theoretical developments, phenomenological studies, and the current status of the Hubble tension, we refer the reader to Refs.~\cite{Poulin:2018cxd,Alexander:2019rsc,Smith:2019ihp,Capparelli:2019rtn,Sakstein:2019fmf,Ye:2020btb,Chudaykin:2020acu,Gogoi:2020qif,Braglia:2020bym,Niedermann:2020dwg,Ivanov:2020ril,DAmico:2020ods,Weiner:2020sxn,Lin:2020jcb,Smith:2020rxx,Murgia:2020ryi,Jedamzik:2020zmd,Chudaykin:2020igl,Fujita:2020ecn,Adi:2020qqf,Seto:2021xua,Tian:2021omz,Freese:2021rjq,DiValentino:2021izs,Sabla:2021nfy,Nojiri:2021dze,Ye:2021nej,Boylan-Kolchin:2021fvy,Vagnozzi:2021gjh,Karwal:2021vpk,Jiang:2021bab,Schoneberg:2021qvd,Poulin:2021bjr,Niedermann:2021ijp,McDonough:2021pdg,LaPosta:2021pgm,Herold:2021ksg,Sabla:2022xzj,Ye:2022afu,Jiang:2022uyg,Rudelius:2022gyu,Berghaus:2022cwf,Ye:2022efx,Reeves:2022aoi,Alexander:2022own,Gomez-Valent:2022bku,Simon:2022adh,Rezazadeh:2022lsf,Smith:2022iax,McDonough:2022pku,Murai:2022zur,Wang:2022bmk,Forconi:2023hsj,Hart:2022agu,Secco:2022kqg,Herold:2022iib,Kamionkowski:2022pkx,Lin:2022phm,Brissenden:2023yko,Reboucas:2023rjm,Cruz:2023cxy,Poulin:2023lkg,Goldstein:2023gnw,Cicoli:2023qri,Jiang:2023bsz,Eskilt:2023nxm,Nojiri:2023mvi,Cruz:2023lmn,Cruz:2023lnq,Liu:2023kce,Peng:2023bik,FrancoAbellan:2023gec,Smith:2023oop,Ramadan:2023ivw,Fu:2023tfo,McDonough:2023qcu,Efstathiou:2023fbn,Gsponer:2023wpm,Khalife:2023qbu,Garny:2024ums,Giare:2024akf,Qu:2024lpx,Seto:2024cgo,Shen:2024hpx,Chatrchyan:2024xjj,Wang:2024tjd,SPT-3G:2025vyw,Carloni:2025jlk,Wang:2025dtk,Andriot:2025los,Poulin:2025nfb,Pang:2025lvh,Allali:2025wwi,Dialektopoulos:2025mfz,Jiang:2025hco,Liu:2026zqc,Bisabr:2026jeo,Yuan:2026xcg,Jhaveri:2026bla,Bella:2026zuk,Schoneberg:2026eys,Schoneberg:2026vaf,Du:2026qtq}} By contrast, purely late-time solutions are strongly restricted~\cite{Knox:2019rjx,Krishnan:2021dyb,Cai:2021weh,Cai:2022dkh,Raveri:2023zmr,Poulin:2024ken,Teixeira:2025czm,Pedrotti:2025ccw,Bansal:2026axl,Pedrotti:2026dwj,Sabogal:2026ipu}. Baryon Acoustic Oscillation (BAO) measurements effectively constrain the Hubble diagram normalization through the combination $H_0\,r_d$, leaving little freedom to raise $H_0$ if the sound horizon is anchored to the $\Lambda$CDM value~\cite{Bernal:2016gxb,Knox:2019rjx,BOSS:2014hhw,Jiang:2025hco,Giare:2026oti}. Moreover, the SH0ES-calibrated supernova (SN) distance ladder and the CMB-BAO inverse distance ladder are themselves in tension. Reconciling this calibration mismatch through late-time modifications to the expansion history alone seems extremely implausible~\cite{Raveri:2023zmr,Poulin:2024ken,Teixeira:2025czm}.\footnote{Uncalibrated SN and BAO tightly constrain the shape of the late-time expansion history~\cite{Efstathiou:2021ocp}. Recent model-independent analyses further limit the late-time contribution to the inferred value of $H_0$ to the percent level, even in the presence of additional early-time physics~\cite{Sabogal:2026ipu}. This argument assumes that new physics does not affect the calibration of the distance ladder itself;~such mechanisms face separate observational constraints, see, e.g., Refs.~\cite{Desmond:2019ygn,Hogas:2025mii,Sakstein:2026grf}.}

Despite these limitations, non-standard late-time dynamics remains independently motivated by recent results from the Dark Energy Spectroscopic Instrument (DESI)~ \cite{DESI:2024mwx,DESI:2025zgx}. The combination of DESI BAO with CMB and SN data shows a preference for departures from a cosmological constant~\cite{DESI:2025fii}, with possible interpretations ranging from an evolving dark energy equation of state (EoS)~\cite{Cortes:2024lgw,Shlivko:2024llw,Luongo:2024fww,Gialamas:2024lyw,Wang:2024dka,Ye:2024ywg,Tada:2024znt,Carloni:2024zpl,Chan-GyungPark:2024mlx,Efstathiou:2024xcq,Bhattacharya:2024hep,Andriot:2024jsh,Li:2024qus,Reboucas:2024smm,Giare:2024gpk,Ramadan:2024kmn,Yang:2024kdo,Jiang:2024xnu,RoyChoudhury:2024wri,Giare:2024oil,Giare:2025pzu,Efstathiou:2025tie,Kessler:2025kju,RoyChoudhury:2025dhe,Scherer:2025esj,Wolf:2025jlc,Santos:2025wiv,Specogna:2025guo,Cheng:2025lod,Cheng:2025hug,Ozulker:2025ehg,Li:2025vuh,Lee:2025pzo,Fazzari:2025lzd,Teixeira:2025czm,Smith:2025icl,Cai:2025mas,Herold:2025hkb,Cheng:2025yue,Chudaykin:2025aux,Chudaykin:2025lww,Chudaykin:2025vdh,Ivanov:2026dvl,Chudaykin:2026nls,Gokcen:2026pkq,Ishak:2025cay,Najafi:2026kxs,Yang:2026yaq,Dinda:2026ktu,Dainotti:2026enb,Kessler:2026dbi,Li:2026xaz,Giare:2026tyk,Li:2026asg,Ren:2026jyw,Andriot:2026lac,Sun:2026kig} to interactions within the dark sector~\cite{Giare:2024smz,Mukhopadhyay:2020bml,Escamilla:2023shf,vanderWesthuizen:2023hcl,Silva:2024ift,DiValentino:2019ffd,Zhao:2022ycr,Li:2024qso,Pooya:2024wsq,Halder:2024uao,Castello:2023zjr,Yao:2023jau,Mishra:2023ueo,Nunes:2016dlj,Silva:2025hxw,Yang:2025uyv,vanderWesthuizen:2025rip,Zhang:2025dwu,Li:2025muv,Li:2025owk,Li:2026xaz,Zhai:2026uwr,Zhang:2026dtd,Sabogal:2026bsw,Sabogal:2024yha,Lewis:2024cqj,Carrion:2025bbk,Zhang:2025lam,Yashiki:2025loj,Tsedrik:2025cwc,Sabogal:2025mkp,Figueruelo:2026eis,Escobal:2026zxb,Dai:2026pvx,Kashyap:2026ivg,Kolhatkar:2026ixl,Antusch:2026ldp,Li:2026asg,Li:2026jgg,Bagherian:2026jxz}.\footnote{Part of this signal can also be accommodated within early-time scenarios~\cite{Chaussidon:2025npr}. EDE can alleviate the DESI-CMB tension~\cite{Poulin:2025nfb}, while dark acoustic oscillations in the early Universe can provide a competitive alternative explanation~\cite{Garny:2025szk}.} Standard evolving dark energy models, such as the Chevallier-Polarski-Linder (CPL) parametrization~\cite{Chevallier:2000qy,Linder:2002et}, tend to worsen the Hubble tension~\cite{DESI:2025fii,Giare:2026oti}. Conversely, certain interacting dark energy (IDE) models can raise $H_0$ in the required direction while preserving a good fit to observations~\cite{Giare:2024smz}. Such late-time expansion-history models, however, remain subject to the general limitations of purely late-time solutions and cannot fully resolve the Hubble tension unless complemented by an early-time mechanism.

This apparent division of labor motivates scenarios in which the dark sector departs from $\Lambda$CDM at both early and late times~\cite{Jedamzik:2020zmd}. The early-time sector should account for most of the shift required in $H_0$, while the late-time sector should accommodate the nonstandard dynamics suggested by low-redshift observations, ideally providing additional help in resolving the Hubble tension. The key question then becomes what properties the late-time sector should have to remain compatible with, and possibly reinforce, the early-time solution.\footnote{Answering this question is also important for recent attempts to embed nonstandard early- and late-time dynamics within a common physical framework~\cite{Ramadan:2023ivw,Giare:2026tyk,SevillanoMunoz:2026jgk,Ling:2026awm}.}

Previous studies have explored several combinations of pre- and post-recombination new physics, showing that individually successful mechanisms do not necessarily cooperate once embedded in the same cosmology~\cite{Toda:2024ncp}. More general analyses have emphasized that late-time modifications may nevertheless help relax some of the parameter shifts required by early-time solutions~\cite{Clark:2021hlo,Pang:2025lvh,Jhaveri:2026bla}. Mixed scenarios combining EDE with interactions in the dark sector have also been considered, primarily using the interaction to address other cosmological tensions rather than to further raise $H_0$~\cite{McDonough:2021pdg,Liu:2023wew,Liu:2023rvo,Simon:2024jmu,Yashiki:2025loj}. What remains less clear is whether an interaction that can itself raise $H_0$ can contribute constructively once an efficient early-time mechanism such as EDE is present.

In this work, we address this question within a broad extended cosmology containing distinct dark sector components: an EDE sector that becomes dynamically relevant before recombination; an independent late-time dark energy fluid with its own energy density and EoS that can exchange energy with cold dark matter (CDM); and a cosmological constant $\Lambda$. Keeping the cosmological constant separate from the interacting fluid allows the amount of dark energy participating in the interaction to vary independently of the total late-time dark energy abundance. \looseness=-1

Focusing on a phenomenological interaction capable of increasing $H_0$, we confront this extended dark sector model with current high- and low-redshift observations. By allowing the EDE and late-time interacting components to vary independently, we directly test whether the interacting fluid can provide additional help in relieving the Hubble tension once EDE is present. We find that the data do not favor such a mixed solution. Although a nonzero interaction can further increase $H_0$, the resulting statistical gain is rapidly outweighed by the deterioration of the fit to low-redshift observations, particularly SN. The interaction is then driven toward the non-interacting limit, while the EDE sector remains essentially unchanged. We identify the physical mechanism that prevents successful cooperation between EDE and the interacting dark energy fluid: energy transfer from dark matter to dark energy requires a reduction in the present-day matter abundance to preserve the CMB-constrained early-time physical matter density, pushing it into a region disfavored by supernova observations. We argue that the resulting limitation extends to a broad class of late-time dark sector interactions with the same direction of energy transfer beyond the specific model considered here.\looseness=-1

The paper is organized as follows. In Sec.~\ref{sec.model} we describe the cosmological model. In Sec.~\ref{sec.method} we present the datasets and methodology. In Sec.~\ref{sec.results} we report the results. In Sec.~\ref{sec.discussion} we discuss their implications for mixed solutions to the Hubble tension, both within and beyond the specific model considered here. We summarize our conclusions in Sec.~\ref{sec.conclusion}.

\section{Theoretical Model}
\label{sec.model}

We consider a spatially flat cosmological model containing radiation, baryons, CDM, an EDE component, an interacting dark energy fluid, and a cosmological constant $\Lambda$.

The EDE component is described by a scalar field with an axion-like potential of the commonly used $n=3$ form, which has provided good fits in EDE analyses including local $H_0$ information~\cite{Kamionkowski:2022pkx,Poulin:2025nfb},
\begin{equation}
V(\phi)=m^2f^2
\left[1-\cos\left(\frac{\phi}{f}\right)\right]^3.
\label{eq:ede_potential}
\end{equation}
The field is initially frozen by Hubble friction and becomes dynamically relevant around the epoch preceding recombination. Once it begins to oscillate about the minimum of the potential, its energy density rapidly dilutes as $\rho_{\rm EDE}\propto a^{-9/2}$, making the EDE contribution negligible at late times. This rapid dilution allows the late-time interacting sector to affect the expansion without a significant residual EDE density that would otherwise spoil the matter era.

To isolate the effects of the interaction, in our baseline analysis the interacting dark energy fluid is characterized by a constant EoS $w$. We relax this assumption in Sec.~\ref{sec.discussion} and Appendix~\ref{app:A}. Its energy-momentum exchange with CDM is specified phenomenologically by
\begin{equation}
\nabla_\mu T_c^{\mu\nu}=Q_c^\nu,
\qquad
\nabla_\mu T_{\rm fld}^{\mu\nu}=-Q_c^\nu.
\label{eq:covariant_interaction}
\end{equation}
We adopt an energy-momentum transfer parallel to the CDM four-velocity $u_c^\nu$ and proportional to the interacting-fluid energy density,
\begin{equation}
Q_c^\nu=\xi\,H\,\rho_{\rm fld}\,u_c^\nu.
\label{eq:interaction}
\end{equation}
The corresponding continuity equations are
\begin{equation}
\begin{aligned}
\dot\rho_c+3H\rho_c &= \xi H\rho_{\rm fld},\\
\dot\rho_{\rm fld}+3H(1+w)\rho_{\rm fld}
&= -\xi H\rho_{\rm fld}.
\end{aligned}
\label{eq:continuity}
\end{equation}
With the sign convention adopted above, $\xi<0$ corresponds to energy transfer from CDM to the interacting fluid, whereas $\xi>0$ describes transfer in the opposite direction. Importantly, the cosmological constant and the interacting fluid are kept as independent components. The $\Lambda$ contribution, as determined by spatial flatness, is allowed to take either sign, providing a broader class of late-time backgrounds. This allows the data to determine how much of the late-time dark energy density, if any, participates in the interaction with CDM, without imposing that the entire accelerated expansion is sourced by the interacting sector. The usual EDE cosmology is continuously recovered when the interacting fluid vanishes, as well as in the limit $\xi\to0$ and $w\to-1$, where the fluid becomes equivalent to a cosmological constant up to a redefinition of $\Omega_\Lambda$.

Solving the continuity equations, the Friedmann equation can be written as
\begin{equation}
\frac{H^2(a)}{H_0^2}
=
\Omega_r a^{-4}
+\Omega_{\rm EDE}(a)
+\Omega_{m,0}a^{-3}
+\Omega_{\rm fld}\frac{\xi}{3w+\xi}a^{-3}
+\Omega_{\rm fld}\frac{3w}{3w+\xi}
a^{-3(1+w)-\xi}
+\Omega_\Lambda.
\label{eq:friedmann_original}
\end{equation}
Here $\Omega_r$ denotes the present-day radiation density parameter, while $\Omega_{m,0}$ denotes the total matter density present today, including baryons, CDM, and massive neutrinos.\footnote{
For simplicity, in Eq.~\eqref{eq:friedmann_original} we do not explicitly separate the massive-neutrino contribution. In our numerical analysis, however, we consistently evolve the neutrino energy density and perturbations, accounting for their transition from radiation-like behavior at early times to matter-like behavior at late times. We assume one massive neutrino with a fixed total mass of $\sum m_\nu=0.06\,\mathrm{eV}$.
} The quantities $\Omega_{\rm fld}$ and $\Omega_\Lambda$ denote the present-day density parameters of the interacting fluid and the cosmological constant, respectively. We further define
$\Omega_{\rm EDE}(a)\equiv\rho_{\rm EDE}(a)/\rho_{\rm cr,0}$ as the EDE energy density normalized to the present-day critical density, $\rho_{\rm cr,0}$.

Equation~\eqref{eq:friedmann_original} shows that the contribution of the interacting fluid can be written as two terms. One scales as ordinary non-relativistic matter, while the other retains a fluid-like evolution. The original parameters $\Omega_{\rm fld}$, $w$, and $\xi$ therefore enter the background expansion only through specific combinations and can exhibit strong degeneracies. It is therefore useful to reorganize the model in terms of \textit{effective} quantities that directly characterize the physical contributions in $H(a)$. We define
\begin{equation}
\Omega_{m,\rm early}^{\rm eff}
\equiv
\Omega_{m,0}+\Delta\Omega_m^{\rm IDE},
\label{eq:omega_m_early}
\end{equation}
where
\begin{equation}
\Delta\Omega_m^{\rm IDE}
\equiv
\Omega_{\rm fld}\frac{\xi}{3w+\xi}.
\label{eq:delta_omega_m}
\end{equation}
The quantity $\Omega_{m,\rm early}^{\rm eff}$ is the amplitude of the total contribution scaling as $a^{-3}$. It therefore corresponds to the effective matter abundance relevant at early times. The CMB will primarily constrain the associated physical density $\omega_{m,\rm early}^{\rm eff} \equiv h^2\, \Omega_{m,\rm early}^{\rm eff}$. By contrast, $\Omega_{m,0}$ is the matter abundance actually present today. Their difference, $\Delta\Omega_m^{\rm IDE}$, quantifies the net change in the matter-like contribution induced by the energy exchange between CDM and the interacting fluid between the early Universe and today.

For the effective fluid, we define
\begin{equation}
w_{\rm eff}
\equiv
w+\frac{\xi}{3},
\qquad
\Omega_{\rm fld}^{\rm eff}
\equiv
\Omega_{\rm fld}\frac{3w}{3w+\xi}.
\label{eq:fluid_eff}
\end{equation}
The effective quantity $w_{\rm eff}$ characterizes the background scaling of this contribution and is not the physical EoS $w$ of the interacting fluid. With these definitions, the Friedmann equation becomes
\begin{equation}
\frac{H^2(a)}{H_0^2}
=
\Omega_r a^{-4}
+\Omega_{\rm EDE}(a)
+\Omega_{m,\rm early}^{\rm eff}a^{-3}
+\Omega_{\rm fld}^{\rm eff}a^{-3(1+w_{\rm eff})}
+\Omega_\Lambda.
\label{eq:friedmann_effective}
\end{equation}

Throughout most of this work, we fix $w\simeq-1$\footnote{To avoid numerical instabilities in the evolution of cosmological perturbations, we fix the equation of state of the interacting fluid to $w=-1+\epsilon$, with $\epsilon=10^{-3}$.} and restrict our analysis to the branch $\xi\le0$, which is the branch where interacting dark energy solutions can raise $H_0$~\cite{Salvatelli:2013wra,Kumar:2016zpg,Murgia:2016ccp,DiValentino:2017iww,DiValentino:2020vnx,Zhai:2023yny,Giare:2024ytc,Giare:2024smz}. Since we also impose $\Omega_{\rm fld}\geq0$, this implies
$\Delta\Omega_m^{\rm IDE}\geq0$. The usual EDE cosmology is continuously recovered when $\Omega_{\rm fld}\rightarrow0$, as well as when $\xi\rightarrow0$ and $w\rightarrow-1$. In the latter case, the interacting fluid becomes degenerate with $\Lambda$ at the background level. Away from these limits, keeping the two components separate allows us to probe a wider range of background dynamics and to generalize our results beyond the $w\simeq-1$ baseline in Sec.~\ref{sec.discussion}.

The effective decomposition introduced above is a background-level reparametrization, while perturbations are evolved in the full model rather than as separately conserved effective components. EDE perturbations follow from the EDE scalar-field dynamics, whereas those of the phenomenological interacting fluid are evolved consistently with the energy transfer specified by $Q_c^\nu\propto u_c^\nu$. This choice implies no momentum transfer in the CDM rest frame. We assume a unit rest-frame sound speed for the interacting fluid, $c_{s,\rm fld}^2=1$, a standard phenomenological choice motivated by the rest-frame sound speed of a canonical scalar field. For further details on the perturbative treatment of interacting dark energy,
we refer the reader to Refs.~\cite{Murgia:2016ccp,DiValentino:2017iww,
Yang:2018uae,Lucca:2020zjb,Giare:2024ytc}.

\section{Data Analysis Methodology}
\label{sec.method}

We implement the cosmological model described in Sec.~\ref{sec.model} in a tailored version of the Boltzmann solver \texttt{CLASS}~\cite{Lesgourgues:2011re,Blas:2011rf}, evolving both the background and perturbations.  

We explore the model parameter space through Markov chain Monte Carlo (MCMC) sampling with \texttt{Cobaya}~\cite{Torrado:2020dgo}. We vary the six standard cosmological parameters (i.e., $\omega_b$, $\omega_c$, $100\theta_s$, $\tau_{\rm reio}$, $\ln(10^{10}A_s)$, and $n_s$), together with the EDE parameters $f_{\rm EDE}$, $\log_{10}z_c$, and $\theta_i$, which specify the maximum fractional energy density of the scalar field, the redshift at which this maximum is reached, and its initial misalignment angle, respectively. When the interacting dark energy sector is active, we sample the native parameters $\Omega_{\rm fld}$ and $\xi$, corresponding to the present-day fluid density and its coupling strength to CDM. The cosmological-constant density $\Omega_\Lambda$ is determined through a shooting procedure enforcing spatial flatness. The uniform prior ranges adopted for the sampled cosmological parameters are listed in Table~\ref{tab:priors}. We require the chains to satisfy the Gelman-Rubin convergence criterion, $R-1\lesssim0.04$.

\begin{table}[htpb!]
\centering
\renewcommand{\arraystretch}{1.15}
\begin{tabular}{l@{\hspace{2cm}}c}
\toprule
Parameter & Prior \\
\midrule
$\omega_b$                 & $[0.005,\,0.1]$ \\
$\omega_{\rm cdm}$         & $[0.001,\,0.99]$ \\
$100\,\theta_s$            & $[0.5,\,10]$ \\
$\tau_{\rm reio}$          & $[0.01,\,0.8]$ \\
$\ln(10^{10}A_s)$          & $[1.61,\,3.91]$ \\
$n_s$                     & $[0.8,\,1.2]$ \\
$f_{\rm EDE}$              & $[0.001,\,0.5]$ \\
$\log_{10}z_c$             & $[3.0,\,4.3]$ \\
$\theta_i$                & $[0.1,\,3.1]$ \\
$\Omega_{\rm fld}$         & $[0,\,1.0]$ \\
$\xi$                     & $[-1.0,\, -10^{-4}]$ \\
\bottomrule
\end{tabular}
\caption{Uniform priors on the sampled cosmological parameters.}
\label{tab:priors}
\end{table}

\clearpage

Our analysis combines the following cosmological datasets:
\begin{itemize}
\item \textbf{CMB:} our CMB dataset combines the following likelihoods and experiments
\begin{itemize}
\item Planck 2018 low-multipole temperature \texttt{Commander} likelihood~\cite{Planck:2019nip}, using TT measurements at $\ell<30$.
\item Planck 2018 low-multipole E-mode polarization likelihood based on \texttt{SRoll2}~\cite{Delouis:2019bub}, using EE measurements at $\ell<30$.
\item Planck 2018 temperature and polarization likelihood~\cite{Planck:2019nip}, restricted to $\ell_{\max}=(1000,600,600)$ for TT, TE, and EE, respectively.
\item Atacama Cosmology Telescope (ACT) DR6 temperature and polarization measurements~\cite{AtacamaCosmologyTelescope:2025blo}, using the TT, TE, and EE spectra at multipoles $\ell\geq600$.
\item South Pole Telescope (SPT)-3G D1 temperature and polarization measurements~\cite{SPT-3G:2025bzu}, using multipoles $\ell\geq400$, with $\ell_{\max}=3000$ for TT and $\ell_{\max}=4000$ for TE and EE.
\item A combined CMB lensing dataset including ACT DR6 lensing~\cite{ACT:2023dou,ACT:2023kun}, Planck lensing~\cite{Carron:2022eyg}, and the SPT-3G \texttt{MUSE} lensing likelihood~\cite{SPT-3G:2024atg}.
\item BICEP/Keck 2018 (BK18) large-scale B-mode polarization likelihood~\cite{BICEP:2021xfz}, based on BB measurements over $20\lesssim\ell\lesssim330$.
\end{itemize}
\item \textbf{DESI:} BAO distance measurements from galaxies, quasars, and the Ly$\alpha$ forest~\cite{DESI:2025qqy,DESI:2025zpo}, including the full covariance between measurements~\cite{DESI:2025zgx}.
\item \textbf{DES-Dovekie (DD):} recalibrated DES 5-year Type Ia supernova likelihood~\cite{DES:2025sig}, based on the Dovekie cross-calibration program and comprising approximately 1600 likely DES supernovae and about 200 low-redshift supernovae from external surveys~\cite{DES:2024jxu,DES:2024upw,DES:2024hip,DES:2025sig}.
\item \textbf{SH0ES}: a Gaussian likelihood on the Hubble constant $H_0=73\pm1$ km/s/Mpc from the SH0ES collaboration~\cite{Riess:2021jrx}, included only in selected dataset combinations.
\end{itemize}

Since the model contains 11 free cosmological parameters, strong degeneracies arise, particularly within the interacting dark sector, where different combinations of the native parameters can produce nearly identical background evolutions. Marginalized posteriors alone may therefore not fully characterize the relevant regions of parameter space, as volume effects can obscure the best-fitting solutions and the associated likelihood cost. For this reason, we complement the MCMC analysis with profile likelihoods for the key parameters.

For a parameter $x$, we define the profile likelihood by minimizing the total $\chi^2$ over all remaining cosmological and nuisance parameters at fixed $x$, and report
\begin{equation}
\Delta\chi^2_{\rm prof}(x)
\equiv
\chi^2_{\rm prof}(x)-\chi^2_{\rm min}.
\end{equation}
Here $\chi^2_{\rm min}$ is the best-fit value obtained when $x$ is also allowed to vary. Both $\chi^2_{\rm prof}(x)$ and $\chi^2_{\rm min}$ are obtained through dedicated likelihood minimizations rather than from the MCMC chains. This provides a volume-independent measure of the relative fit quality and allows us to identify both global and local best-fitting regions.

\section{Results}
\label{sec.results}

\begin{figure}[htpb!]
    \centering
    \includegraphics[width=\linewidth]{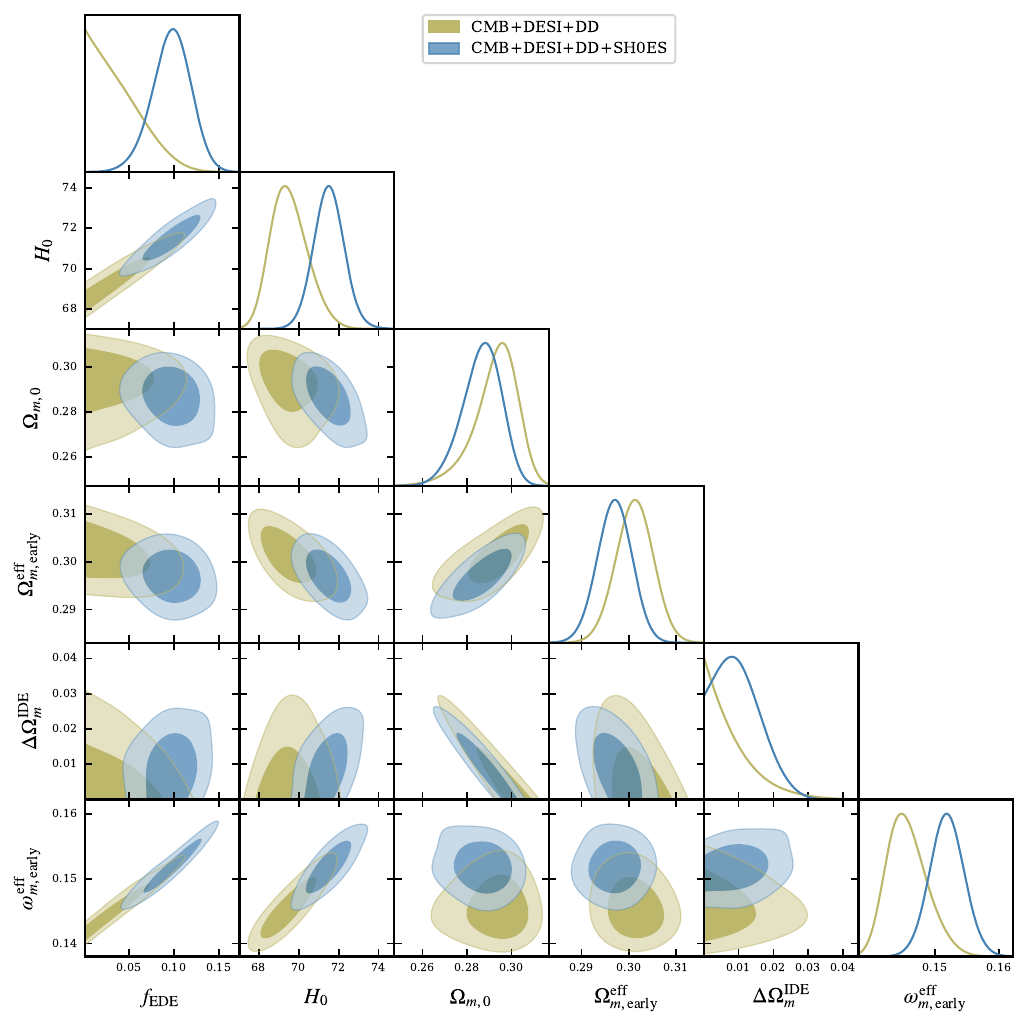}
    \caption{One- and two-dimensional marginalized posterior distributions for a representative subset of parameters in the EDE+IDE model, obtained from CMB+DESI+DD and CMB+DESI+DD+SH0ES. The two-dimensional contours show the $68\%$ and $95\%$ CL regions.}
    \label{fig:t_plot}
\end{figure}

Table~\ref{tab:mcmc_constraints} summarizes the
constraints on the cosmological parameters for CMB+DESI+DD and CMB+DESI+DD+SH0ES, while Figure~\ref{fig:t_plot} shows the corresponding one- and two-dimensional marginalized posteriors for selected parameters. Unless otherwise stated, we quote marginalized constraints at 68\% confidence level (CL) and upper limits at 95\% CL. 

\begin{table*}[htpb!]
\centering
\renewcommand{\arraystretch}{1.4}
\resizebox{0.75\textwidth}{!}{
\begin{tabular}{l c c}
\toprule
Parameter
& CMB+DESI+DD
& CMB+DESI+DD+SH0ES \\
\midrule

$100\,\omega_b$
& $2.253 \pm 0.013$
& $2.265 \pm 0.012$ \\

$\omega_{\rm cdm,0}$
& $0.1185^{+0.0049}_{-0.0038}$
& $0.1233 \pm 0.0038$ \\

$100\,\theta_s$
& $1.04158 \pm 0.00027$
& $1.04124 \pm 0.00028$ \\

$\tau_{\rm reio}$
& $0.0633 \pm 0.0059$
& $0.0616^{+0.0046}_{-0.0061}$ \\

$\ln(10^{10}A_s)$
& $3.068 \pm 0.011$
& $3.074 \pm 0.009$ \\

$n_s$
& $0.9778^{+0.0048}_{-0.0061}$
& $0.9859 \pm 0.0043$ \\

\hline

$f_{\rm EDE}$
& $<0.092$
& $0.097 \pm 0.020$ \\

$\log_{10}z_c$
& $3.506^{+0.122}_{-0.167}$
& $3.542 \pm 0.043$ \\

$\theta_i$
& $>0.38$
& $2.68^{+0.17}_{-0.05}$ \\

\hline

$\Delta\Omega_m^{\rm IDE}$
& $<0.0230$
& $<0.0217$ \\

$\Omega_{m,\rm early}^{\rm eff}$
& $0.3012 \pm 0.0038$
& $0.2970 \pm 0.0035$ \\

$\omega_{m,\rm early}^{\rm eff}$
& $0.1456^{+0.0019}_{-0.0038}$
& $0.1518 \pm 0.0026$ \\

$w_{\rm eff}$
& $-1.083^{+0.086}_{-0.017}$
& $-1.013^{+0.010}_{-0.007}$ \\

\hline

$H_0\ [{\rm km\,s^{-1}Mpc^{-1}}]$
& $69.51^{+0.65}_{-1.08}$
& $71.51 \pm 0.71$ \\

$r_{\rm drag}\ [{\rm Mpc}]$
& $145.24^{+2.04}_{-0.92}$
& $141.96 \pm 1.19$ \\

$\Omega_{m,0}$
& $0.2932^{+0.0109}_{-0.0047}$
& $0.2867^{+0.0095}_{-0.0075}$ \\

$\sigma_8$
& $0.850^{+0.010}_{-0.028}$
& $0.870^{+0.019}_{-0.028}$ \\

$S_8$
& $0.840^{+0.009}_{-0.014}$
& $0.850^{+0.011}_{-0.014}$ \\

\bottomrule
\end{tabular}
}
\caption{Marginalized constraints on the cosmological parameters of the EDE+IDE model for CMB+DESI+DD and CMB+DESI+DD+SH0ES. We report marginalized constraints at $68\%$ CL and upper limits at $95\%$ CL.}
\label{tab:mcmc_constraints}
\end{table*}

The main effect of including SH0ES is to select a non-zero EDE component. The constraint on $f_{\rm EDE}$ shifts from $f_{\rm EDE}<0.092$ for CMB+DESI+DD to $f_{\rm EDE}=0.097\pm0.020$ for CMB+DESI+DD+SH0ES. Correspondingly, $H_0$ increases from $69.51^{+0.65}_{-1.08}$ km/s/Mpc to $71.51\pm0.71$ km/s/Mpc. 

By contrast, the late-time dark sector is largely unaffected by the inclusion of SH0ES. While the native parameters remain strongly degenerate, the effective quantities introduced in Sec.~\ref{sec.model} isolate the physically distinct contributions to the expansion history and are therefore better constrained by the data. For both datasets, we find only upper limits on the net amount of energy transferred from CDM to the late-time dark-energy fluid: $\Delta\Omega_m^{\rm IDE}<0.0230$ for CMB+DESI+DD and $\Delta\Omega_m^{\rm IDE}<0.0217$ for CMB+DESI+DD+SH0ES. The effective early-time matter abundance also changes only mildly, from $\Omega_{m,\rm early}^{\rm eff}=0.3012\pm0.0038$ to $\Omega_{m,\rm early}^{\rm eff}=0.2970\pm0.0035$. These constraints already indicate that the local $H_0$ information drives the model toward a non-zero EDE contribution, but not toward a non-zero dark-sector interaction.

\begin{table*}[t!]
\centering
\renewcommand{\arraystretch}{1.2}
\resizebox{0.8\textwidth}{!}{%
\begin{tabular}{lcccc}
\toprule
&
\multicolumn{2}{c}{CMB+DESI+DD}
&
\multicolumn{2}{c}{CMB+DESI+DD+SH0ES}
\\
\cmidrule(lr){2-3}
\cmidrule(lr){4-5}
Parameter
& EDE-only
& EDE+IDE
& EDE-only
& EDE+IDE
\\
\midrule

$100\,\omega_b$
& $2.2591$
& $2.2584$
& $2.2599$
& $2.2578$
\\

$\omega_{\rm cdm,0}$
& $0.1273$
& $0.1273$
& $0.1274$
& $0.1274$
\\

$100\,\theta_s$
& $1.04127$
& $1.04124$
& $1.04128$
& $1.04125$
\\

$\tau_{\rm reio}$
& $0.0614$
& $0.0614$
& $0.0614$
& $0.0612$
\\

$\ln(10^{10}A_s)$
& $3.0715$
& $3.0715$
& $3.0716$
& $3.0716$
\\

$n_s$
& $0.9823$
& $0.9821$
& $0.9825$
& $0.9822$
\\

\hline
$f_{\rm EDE}$
& $0.0867$
& $0.0863$
& $0.0904$
& $0.0904$
\\

$\log_{10}z_c$
& $3.533$
& $3.535$
& $3.526$
& $3.521$
\\

$\theta_i$
& $2.716$
& $2.733$
& $2.750$
& $2.765$
\\
\hline
$\xi$
& --
& $-1.2\times10^{-4}$
& --
& $-3.5\times10^{-4}$
\\

$\Delta\Omega_m^{\rm IDE}$
& --
& $2.41\times10^{-5}$
& --
& $8.44\times10^{-5}$
\\

$\Omega_{m,\rm early}^{\rm eff}$
& --
& $0.3022$
& --
& $0.2998$
\\

$\omega_{m,\rm early}^{\rm eff}$
& --
& $0.1505$
& --
& $0.1506$
\\

$w_{\rm eff}$
& --
& $-0.999$
& --
& $-0.999$
\\

\hline
$H_0\ [{\rm km\,s^{-1}Mpc^{-1}}]$
& $70.63$
& $70.57$
& $70.92$
& $70.88$
\\
$r_{\rm drag}\ [{\rm Mpc}]$
& $142.63$
& $142.65$
& $142.47$
& $142.49$
\\

$\Omega_{m,0}$
& $0.3018$
& $0.3022$
& $0.2996$
& $0.2997$
\\

$\sigma_8$
& $0.8332$
& $0.8329$
& $0.8328$
& $0.8326$
\\

$S_8$
& $0.8357$
& $0.8360$
& $0.8322$
& $0.8322$
\\
\addlinespace

\hline 
$\Delta\chi^2_{\rm tot}$
& \multicolumn{2}{c}{$-0.16$}
& \multicolumn{2}{c}{$-0.02$}
\\

\bottomrule
\end{tabular}%
}
\caption{Best-fit cosmological parameters for the EDE-only and EDE+IDE models for the two dataset combinations considered. The last row reports $\Delta\chi^2_{\rm tot}\equiv\chi^2_{\rm tot}(\mathrm{EDE+IDE})-\chi^2_{\rm tot}(\mathrm{EDE\mbox{-}only})$, where $\chi^2_{\rm tot}$ includes the calibration-prior contributions.}
\label{tab:profile_best_fits}
\end{table*}

This indication is confirmed at the likelihood level. We perform dedicated likelihood minimizations for both the EDE-only model, in which the interacting fluid is switched off, and the full EDE+IDE model. Table~\ref{tab:profile_best_fits} compares the corresponding best-fit parameters for both dataset combinations. The two solutions are essentially identical: the EDE parameters remain unchanged, while the interaction approaches the non-interacting limit, with $\xi\sim10^{-4}$ for both datasets. Despite two additional cosmological parameters, the EDE+IDE model improves the fit by only $\Delta\chi^2_{\rm tot}=-0.16$ and $-0.02$, respectively. The best-fit solution therefore collapses onto the EDE-only model.

\begin{figure}[htpb!]
    \centering
    \includegraphics[width=0.8\linewidth]{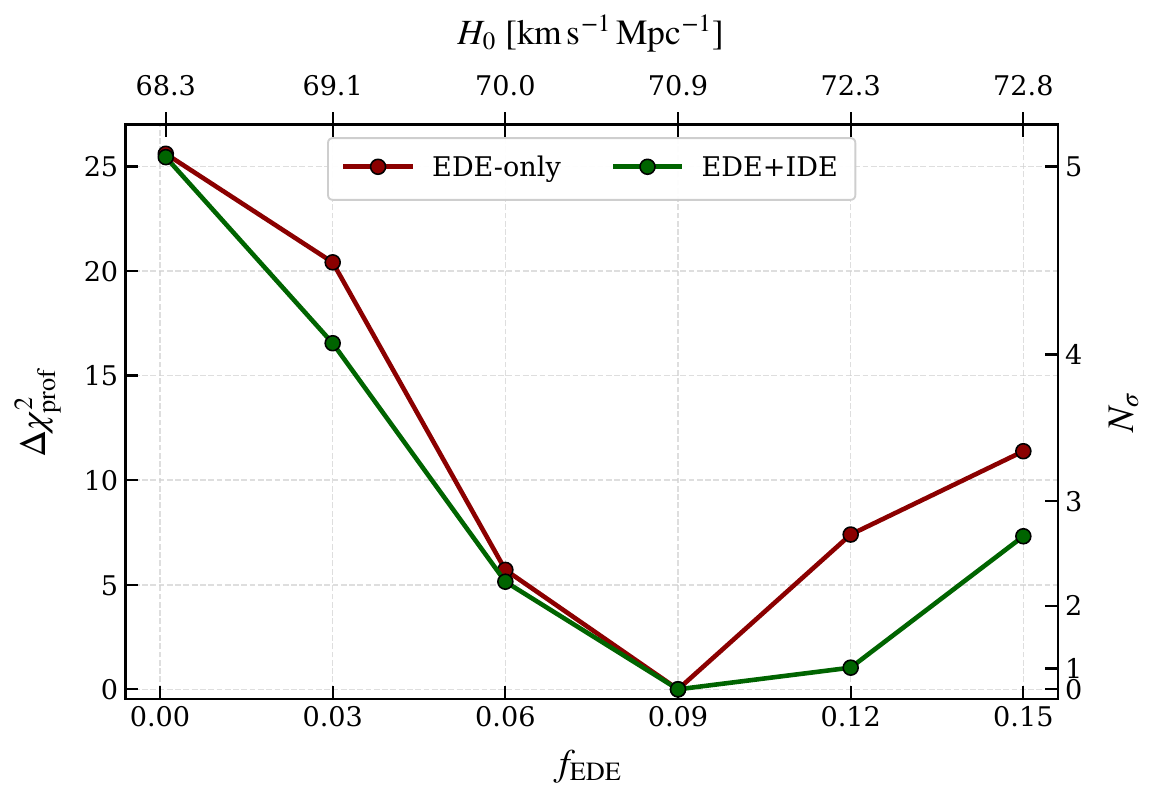}
    \caption{Profile likelihood of $f_{\rm EDE}$ for the EDE-only and EDE+IDE models using CMB+DESI+DD+SH0ES. The lower horizontal axis shows $f_{\rm EDE}$, while the upper horizontal axis gives the corresponding value of $H_0$, which is nearly identical for the two models along the profile. The left vertical axis shows $\Delta\chi^2_{\rm prof}$ relative to the global minimum, while the right vertical axis gives the corresponding significance, $N_\sigma=\sqrt{\Delta\chi^2_{\rm prof}}$.}
    \label{fig:fede_profile}
\end{figure}

Figure~\ref{fig:fede_profile} shows the profile likelihood of $f_{\rm EDE}$ for both the EDE-only and EDE+IDE models using CMB+DESI+DD+SH0ES. The two profiles have identical minima at $f_{\rm EDE}\simeq0.09$, showing that the preferred EDE solution is unchanged by the interacting sector. Along the profile, the corresponding values of $H_0$ are also nearly identical in the two models (top horizontal axis), so that the interaction does not provide an additional shift in the Hubble constant. Its only appreciable effect is to reduce $\Delta\chi^2_{\rm prof}$ away from the minimum, particularly toward larger $f_{\rm EDE}$ and $H_0$, where the additional freedom partially compensates the worsening fit.

\begin{figure}[htpb!]
    \centering
    \includegraphics[width=0.8\linewidth]{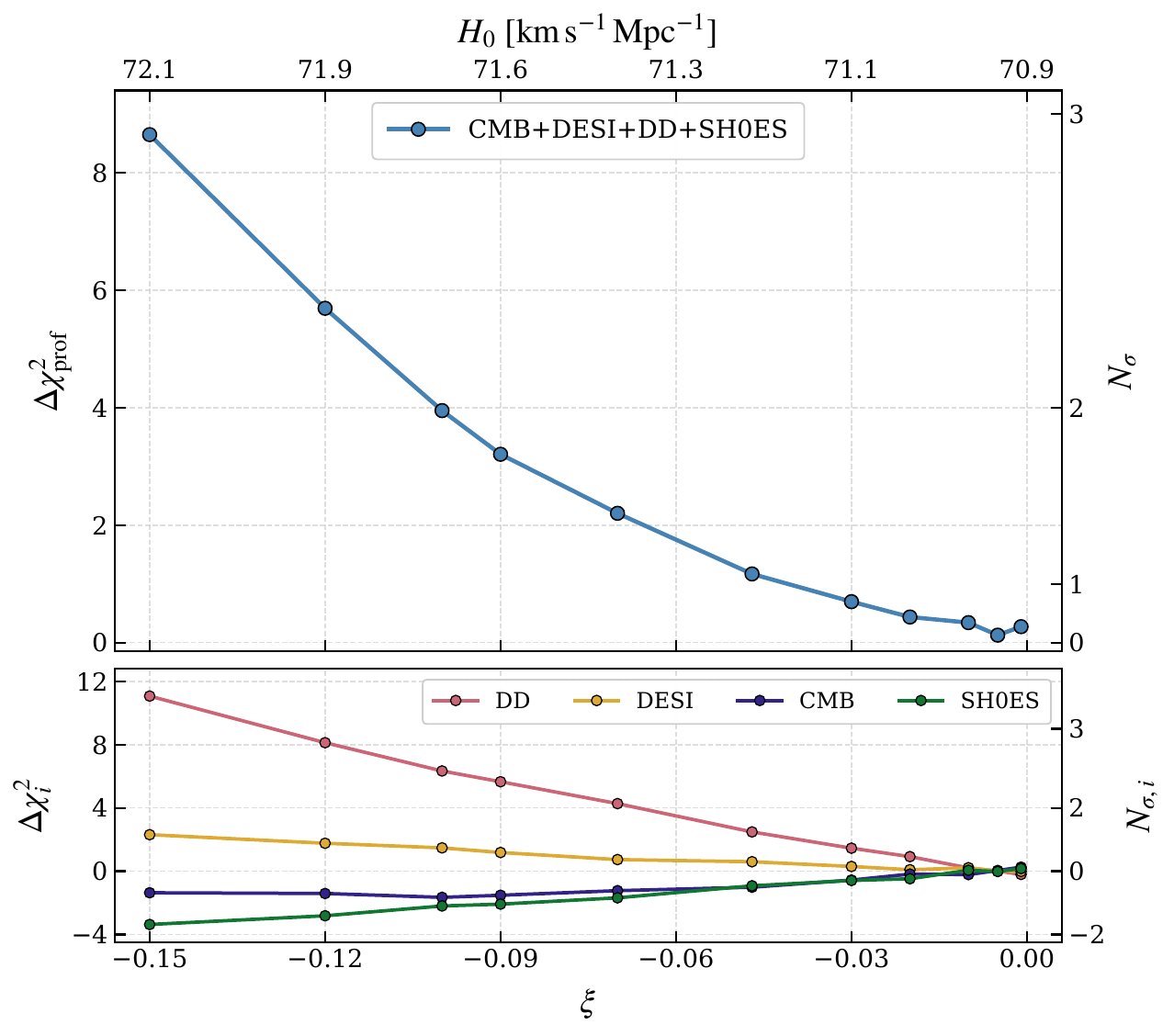}
    \caption{Profile likelihood of the interaction strength $\xi$ for the EDE+IDE model keeping $\Omega_\Lambda=0$ fixed and using CMB+DESI+DD+SH0ES. In the upper panel, the lower horizontal axis shows $\xi$, while the upper horizontal axis gives the corresponding value of $H_0$ along the profile. The left vertical axis shows $\Delta\chi^2_{\rm prof}$ relative to the global minimum, while the right vertical axis gives the corresponding significance, $N_\sigma=\sqrt{\Delta\chi^2_{\rm prof}}$. The lower panel shows the individual contributions $\Delta\chi_i^2$ from DD, DESI, CMB, and SH0ES along the same profile. The right vertical axis gives the corresponding signed significance, $N_{\sigma,i}=\mathrm{sgn}(\Delta\chi_i^2)\sqrt{|\Delta\chi_i^2|}$, so that positive and negative values indicate a worsening or improvement of the individual likelihood, respectively.}
    \label{fig:xi_profile}
\end{figure}

To understand why the interacting dark sector fails to cooperate with EDE in raising $H_0$, we profile the interaction strength itself. For this specific test we fix $\Omega_{\Lambda}=0$ and assume that late-time accelerated expansion is entirely driven by the interacting fluid.\footnote{For the $\xi$ profile, we fix $\Omega_\Lambda=0$ and determine $\Omega_{\rm fld}$ through a shooting procedure enforcing spatial flatness at each fixed $\xi$. If $\Omega_\Lambda$ were allowed to vary, the limit $\Omega_{\rm fld}\to0$ would recover the standard EDE-only model independently of $\xi$, making its best-fit $\chi^2$ accessible for any value of the interaction. We have explicitly verified that the EDE-only limit and the $\Omega_\Lambda=0$ limit with $\xi\to0$ yield essentially identical best-fit $\chi^2$, while no other configuration explored provides a significant improvement. Therefore, fixing $\Omega_\Lambda=0$ removes this degeneracy without excluding any significantly better-fitting solution within the parameter space considered.} Figure~\ref{fig:xi_profile} shows the profile likelihood of $\xi$ for CMB+DESI+DD+SH0ES, with the corresponding value of $H_0$ shown on the upper horizontal axis. The global minimum lies close to $\xi\simeq0$, corresponding to the non-interacting limit. Moving toward increasingly negative $\xi$ raises $H_0$, improving the SH0ES fit. The CMB contribution also decreases mildly, favoring the same direction. DESI remains largely neutral for small interactions and only mildly disfavors larger negative $\xi$. The dominant penalty instead comes from SN: the DD $\chi^2$ worsens rapidly with the interaction strength, overwhelming the improvements from SH0ES and the CMB.

The competition between the SH0ES gain and the SN penalty can be quantified directly. The EDE-only best fit to the same dataset combination gives $H_0=70.92$ km/s/Mpc. Since the SH0ES likelihood is taken to be Gaussian with $H_0=73\pm1$ km/s/Mpc, its residual contribution at the EDE-only best fit is $\chi^2_{\rm SH0ES}\simeq 4.3$. Thus, even in the ideal case in which the interaction shifted $H_0$ exactly to the SH0ES central value, the maximum improvement available from SH0ES would be only $\Delta\chi^2_{\rm SH0ES}\simeq-4.3$. By contrast, along the $\xi$ profile the DD contribution alone reaches a penalty of approximately $\Delta\chi^2_{\rm DD}\simeq11$ at $\xi\simeq-0.15$, while the total profile likelihood has already worsened by $\Delta\chi^2_{\rm prof}\simeq 8.6$. The SN penalty therefore becomes substantially larger than the entire residual gain available from SH0ES, and the mild improvement in the CMB likelihood is insufficient to compensate for it. 

\begin{figure}[htpb!]
    \centering
    \includegraphics[width=0.8\linewidth]{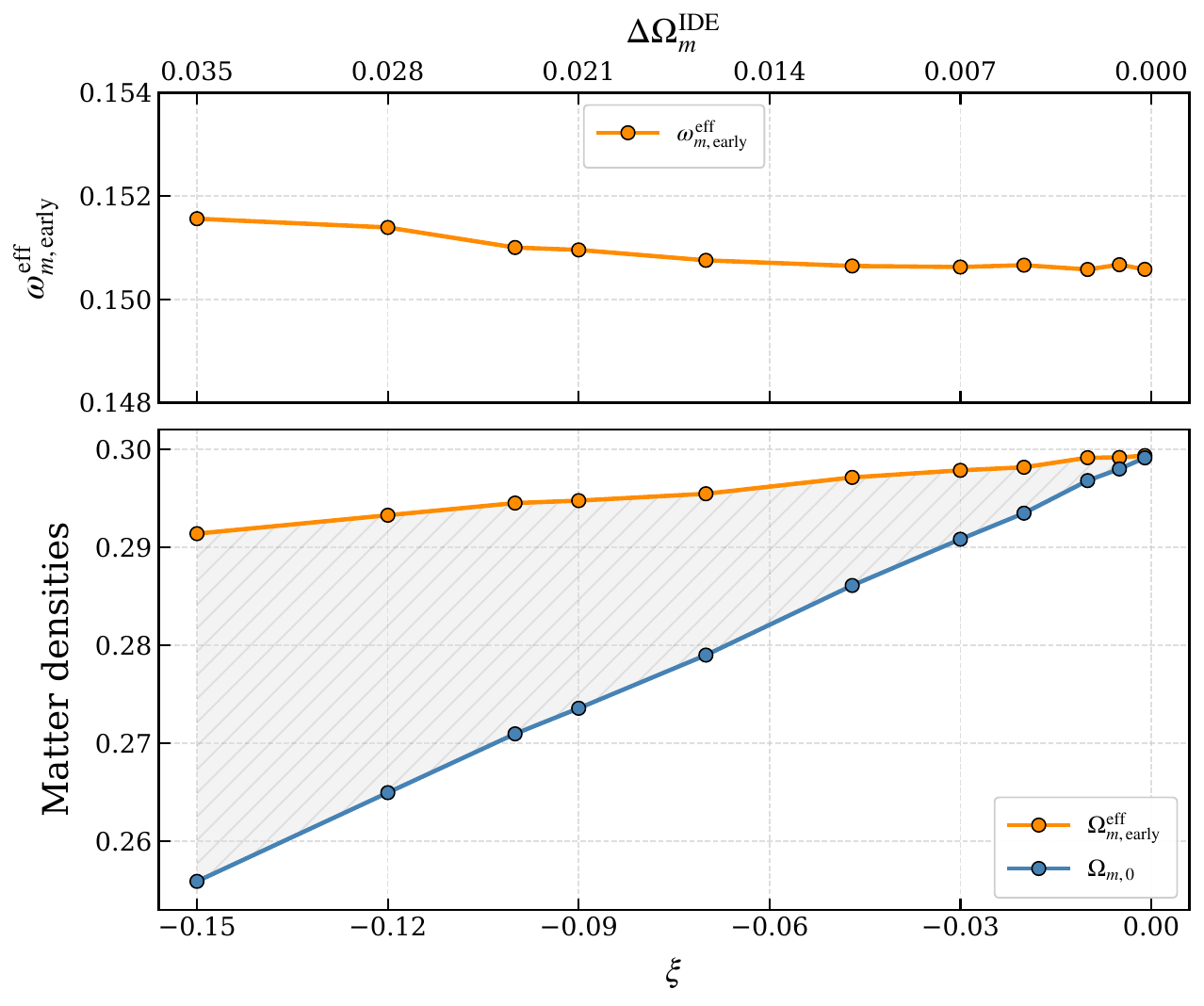}
    \caption{Evolution of the matter densities along the $\xi$ profile for the EDE+IDE model keeping $\Omega_\Lambda=0$ and using CMB+DESI+DD+SH0ES. The upper panel shows the effective early-time physical matter density $\omega_{m,\rm early}^{\rm eff}$, while the lower panel shows the effective early-time matter density $\Omega_{m,\rm early}^{\rm eff}$ and the present-day matter density $\Omega_{m,0}$. The lower horizontal axis gives $\xi$, while the upper horizontal axis shows the corresponding $\Delta\Omega_m^{\rm IDE}$, which quantifies the net amount of energy transferred from CDM to the interacting dark energy fluid.}
    \label{fig:matter_profile}
\end{figure}

The mechanism by which the SN data suppress the interaction is illustrated in Figure~\ref{fig:matter_profile}, which shows the evolution of the matter densities along the $\xi$ profile. The effective early-time physical matter density $\omega_{m,\rm early}^{\rm eff}$, given by
\begin{equation}
\omega_{m,\rm early}^{\rm eff}
=
h^2\left(\Omega_{m,0}+\Delta\Omega_m^{\rm IDE}\right),
\label{eq:failure}
\end{equation}
remains essentially constant across the entire profile, varying by less than $0.7\%$. This is expected, since it is precisely the physical matter density probed and tightly constrained by the CMB~\cite{Eisenstein:2004an,Hu:2001bc,Hu:2000ti,Hu:1995kot,Hu:2004kn,Ivanov:2020mfr,Planck:2018vyg,Loverde:2024nfi,Costa:2025kwt,Weiner:2026sfm,Pedrotti:2026ncd}. Instead, moving toward more negative $\xi$ increases both $\Delta\Omega_m^{\rm IDE}$ and $H_0$, and hence $h^2$. Keeping $\omega_{m,\rm early}^{\rm eff}$ nearly fixed therefore requires $\Omega_{m,0}$ to decrease, as seen in Figure~\ref{fig:matter_profile}. This is precisely the direction strongly penalized by the SN data, which tightly constrain $\Omega_{m,0}$ through the low-redshift distance-redshift relation~\cite{Brout:2022vxf,DES:2024jxu,DES:2024hip,DES:2025sig}. The combination of CMB and SN therefore strongly restricts any nonzero $\Delta\Omega_m^{\rm IDE}$ and drives the interaction toward zero.

Importantly, this mechanism already represents a restriction in IDE-only models (i.e., without EDE), where an interaction-induced increase in $H_0$ must be accompanied by a decrease in $\Omega_{m,0}$ to keep $\omega_{m,\rm early}^{\rm eff}$ fixed. In that case, however, the interaction acts on the lower $H_0$ associated with an otherwise standard early-time cosmology. As a result, a moderate increase can be accommodated without immediately pushing $\Omega_{m,0}$ into the region strongly disfavored by low-redshift data. Moreover, the larger initial discrepancy with SH0ES means that increasing $H_0$ yields a larger statistical gain, partially compensating the low-redshift penalty. The presence of EDE makes this balance substantially more restrictive. EDE already raises $H_0$ independently of the interaction, increasing $h^2$ before any energy transfer is introduced and absorbing most of the SH0ES gain. Turning on the interaction then further raises $H_0$ while generating a positive $\Delta\Omega_m^{\rm IDE}$. This requires a larger reduction in $\Omega_{m,0}$, while only a limited residual gain from SH0ES remains to offset the low-redshift penalty. The parameter space in which the interaction can provide additional relief to the Hubble tension without incurring a large SN penalty is therefore strongly reduced once EDE is present.

\section{Discussion}
\label{sec.discussion}

The results of the previous section show that, in the model considered here, the interacting dark sector does not cooperate with EDE in raising $H_0$: the CMB tightly constrains the early-time matter density, while SN constrain the present-day matter abundance. We now place this result in the broader context of mixed early- and late-time solutions to the Hubble tension and argue that the same physical mechanism extends to a broader class of late-time dark-sector interactions.

For a mixed solution in which both the early- and late-time expansion histories are modified, keeping the CMB angular scale $\theta_s(z_*)$ fixed gives
$\Delta\ln H_0\simeq-\Delta\ln r_s(z_*)+\Delta\ln I_*$,
where $\Delta\ln r_s(z_*)$ is the fractional variation of the sound horizon at recombination and $\Delta\ln I_*$ is the fractional variation of the dimensionless comoving-distance integral
\begin{equation}
I_* \equiv \int_0^{z_*}\frac{dz}{E(z)},
\qquad
E(z)\equiv\frac{H(z)}{H_0}.
\label{eq:Istar}
\end{equation}
The early-time contribution raises $H_0$ through $\Delta\ln r_s(z_*)<0$, while any additional late-time contribution requires $\Delta\ln I_*>0$. Taking the first-order variation of Eq.~\eqref{eq:Istar},\footnote{Here we neglect the small model dependence of $z_*$.}
\begin{equation}
\delta\ln I_*
=
-\frac{1}{I_*}
\int_0^{z_*}
\frac{\delta E(z)}{E^2(z)}\,dz ,
\label{eq:delta_I}
\end{equation}
shows that $\Delta\ln I_*>0$ requires a net weighted decrease in the normalized expansion rate $E(z)$ over the redshift range contributing to the distance integral.

\begin{figure}[t!]
\centering
\includegraphics[width=0.9\linewidth]{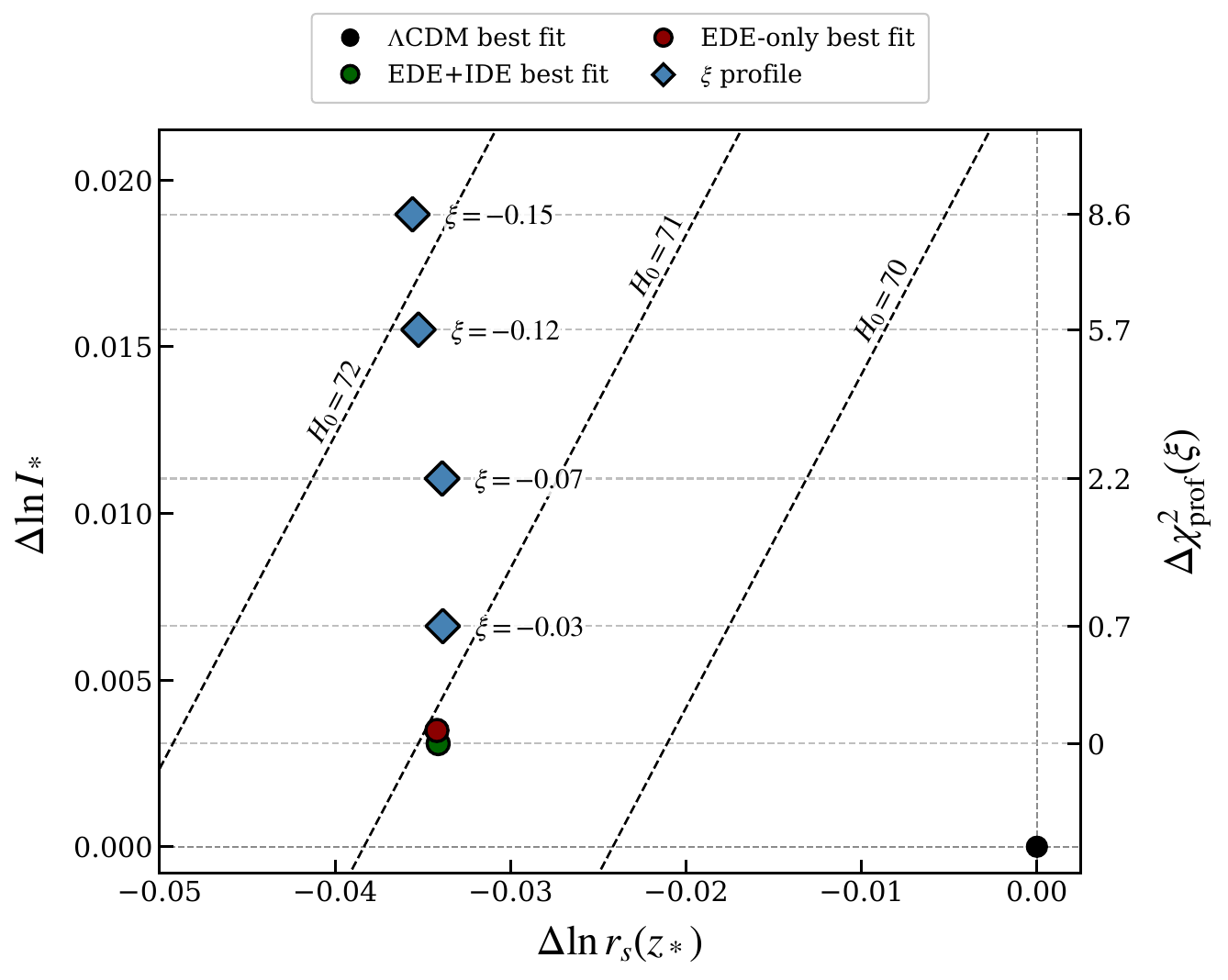}
\caption{Results for CMB+DESI+DD+SH0ES in the physical plane spanned by $\Delta\ln r_s(z_*)$ and $\Delta\ln I_*$. The origin corresponds to the $\Lambda$CDM best fit. The red and green points show the best-fit EDE-only and EDE+IDE models, respectively, while the blue points show selected positions along the $\xi$ profile; the right vertical axis shows the corresponding $\Delta\chi^2_{\rm prof}(\xi)$. The diagonal lines indicate lines of constant $H_0$.}
\label{fig:mixed_solutions}
\end{figure}

Figure~\ref{fig:mixed_solutions} maps our results onto the physical plane spanned by $\Delta\ln r_s(z_*)$ and $\Delta\ln I_*$. By construction, the origin corresponds to the $\Lambda$CDM best fit. The EDE-only and EDE+IDE best fits for CMB+DESI+DD+SH0ES (red and green points, respectively) produce essentially identical shifts relative to $\Lambda$CDM, dominated in both cases by a negative $\Delta\ln r_s(z_*)$. This reflects the standard EDE mechanism, in which the enhanced pre-recombination expansion rate reduces the sound horizon. The much smaller shift in $\Delta\ln I_*$ is instead due to modest differences in the best-fit late-time cosmological parameters.

The blue points show how the EDE+IDE model moves across this plane as the interaction is progressively forced away from zero. As $\xi$ becomes more negative, the model shifts toward positive $\Delta\ln I_*$ and higher $H_0$. A non-vanishing interaction can therefore further raise $H_0$ through $\Delta\ln I_*>0$, but the same trajectory rapidly accumulates a large likelihood penalty, as shown by the increasing $\Delta\chi^2_{\rm prof}$ values on the right-hand vertical axis.

This mechanism is not specific to the interaction adopted here, but extends to a broad class of late-time IDE models beyond our baseline assumptions. Consider a broader class of phenomenological interacting dark energy fluids with equation of state $w(a)$. For convenience, we define $Q^\nu\equiv-Q_c^\nu$, such that $\nabla_\mu T_c^{\mu\nu}=-Q^\nu$ and $\nabla_\mu T_{\rm DE}^{\mu\nu}=Q^\nu$, with $Q^\nu=Q(a)u^\nu$ at the background level. In this convention, energy transfer from CDM to dark energy corresponds to $Q(a)>0$. Since we are interested in genuinely mixed early- and late-time solutions, we restrict this generalization to interactions that become relevant after recombination. For such a late-time interaction, the net amount of energy transferred from CDM between recombination and today can be written as
\begin{equation}
\Delta\Omega_m^Q
\equiv
\frac{1}{\rho_{\rm cr,0}}
\int_{a_*}^{1}
\frac{a^2Q(a)}{H(a)}\,da
>0.
\end{equation}
The effective physical matter density probed by the CMB is then $\omega_{m,\rm early}^{\rm eff}=h^2(\Omega_{m,0}+\Delta\Omega_m^Q)$. For an interaction with $\Delta\Omega_m^Q>0$ that raises $H_0$, keeping $\omega_{m,\rm early}^{\rm eff}$ fixed requires a lower present-day matter abundance $\Omega_{m,0}$. This is precisely the direction strongly constrained by the SN data. Within this class of post-recombination interactions, changing the functional form of $Q(a)$ can modify the timing of the energy transfer and its impact on perturbations, but does not alter the integrated relation between the early-time and present-day matter abundances.

The remaining freedom in the dark energy EoS does not provide an obvious way around this limitation. For the phenomenological interacting fluid models considered here, the coupling can source large-scale non-adiabatic instabilities in the dark energy perturbations. Within the standard fluid treatment, their stability can be characterized by the doom factor $d(a)\equiv-Q(a)/\left[3H\rho_{\rm DE}(1+w(a))\right]$, which enters the coupling-dependent terms governing the large-scale evolution of these perturbations~\cite{Valiviita:2008iv,Gavela:2009cy}. Values $d(a)>0$ generate a rapidly growing non-adiabatic mode, while stability requires $d(a)<0$. 
For energy transfer from CDM to dark energy, $Q(a)>0$, the perturbative stability condition $d(a)<0$ therefore requires $w(a)>-1$ within this phenomenological fluid framework.\footnote{We stress that this condition applies to the physical EoS of the interacting fluid. A phantom-like effective expansion history does not necessarily require a phantom component. Multi-component or interacting dark sectors can mimic $w_{\rm eff}<-1$ when interpreted within a non-interacting single-fluid description, see e.g., Refs.~\cite{Das:2005yj,Chakraborty:2025syu,Liu:2025bss,Khoury:2025txd,Guedezounme:2025wav,Chen:2025ywv,Antusch:2026ldp,Li:2026xaz,Ganguly:2026mgu,Khoury:2026svx}.} At fixed $Q(a)$ and present-day dark energy density, moving away from $w(a)=-1$ into the non-phantom regime increases the dark energy density in the past relative to the vacuum-like limit, thereby increasing $E(z)$. According to Eq.~\eqref{eq:delta_I}, this gives a negative contribution to $\Delta\ln I_*$. This acts in the opposite direction to that required to raise $H_0$ and must therefore be compensated by a stronger interaction to achieve the same increase in $I_*$. The resulting larger energy transfer from CDM to dark energy forces $\Omega_{m,0}$ to decrease further in order to keep $\omega_{m,\rm early}^{\rm eff}$ fixed. Approaching $w(a)=-1$ from the stable branch therefore minimizes this adverse contribution to $\Delta\ln I_*$. Therefore, allowing $w(a)$ to evolve can enlarge the available parameter space through additional degeneracies and may weaken the inferred low-redshift constraints, but it does not provide a more efficient physical mechanism for raising $H_0$, as shown in Ref.~\cite{Giare:2024ytc}.

We have explicitly verified this expectation by allowing the EoS of the late-time dark energy fluid to evolve according to a CPL parametrization. The full results are reported in Appendix~\ref{app:A}. When both the interaction and the CPL parameters are varied,
the best fit lies at the prior boundary near the non-interacting limit, with $\xi=-10^{-4}$, and improves the total fit by only $\Delta\chi^2_{\rm tot}\simeq-2.4$. Within the stable non-phantom branch, the best fit lies very close to the $w(a)=-1$ limit adopted throughout our baseline analysis. This confirms that fixing the fluid EoS close to $w=-1$ does not drive our main result.

The limited improvement in the fit is also a consequence of this restriction. Indeed, when the interaction is switched off and the CPL EoS is varied within the same non-phantom region, we find a comparable improvement, $\Delta\chi^2_{\rm tot}\simeq-2.4$. This shows that the small gain obtained in the EDE+IDE-CPL model is primarily associated with the additional EoS freedom rather than with the interaction itself. If the CPL EoS is instead allowed to explore the full parameter space, the improvement becomes substantially larger, $\Delta\chi^2_{\rm tot}\simeq-8.9$. However, the preferred solution crosses into the phantom regime and lowers $H_0$ to $70.3$ km/s/Mpc. Thus, additional freedom in the dark energy EoS can significantly improve the overall fit, but does not restore a constructive role for the interaction in raising $H_0$. 

Finally, we note that this mechanism is not specific to the EDE potential considered here. It may also limit the additional contribution of dark sector interactions when combined with other early-time solutions that raise $H_0$ primarily by reducing the sound horizon. Provided that the CMB-constrained early-time physical matter density remains nearly fixed as the interaction is varied, any additional energy transfer from CDM to dark energy further lowers the present-day matter abundance, potentially worsening the fit to low-redshift observations. Whether this produces a comparable likelihood penalty, however, remains model dependent and requires explicit verification.

\section{Conclusions}
\label{sec.conclusion}

In this work, we have investigated whether IDE can provide additional help in relieving the Hubble tension once EDE is present. 

We considered an extended dark sector consisting of an EDE component active before recombination, an independent late-time dark energy fluid interacting with CDM, and a cosmological constant. The energy-transfer rate was chosen to be proportional to the energy density of the interacting fluid. For this interaction, the branch corresponding to energy transfer from CDM to dark energy is known to shift $H_0$ toward larger values.

Confronting this model with current high- and low-redshift observations, we find that the data do not favor a late-time IDE sector. The dark energy fluid is pushed toward the non-interacting limit, and the best-fitting cosmology corresponds to the EDE-only limit of the model.

Although IDE can in principle provide an additional increase in $H_0$ toward the value measured by SH0ES, EDE already accounts for most of the shift required to alleviate the Hubble tension. At the EDE-only best fit, the remaining improvement available from the SH0ES likelihood is $\Delta\chi^2\simeq-4$. In the presence of a non-vanishing interaction, this limited gain is rapidly outweighed by the worsening fit to low-redshift observations, particularly SN data, with a degradation of $\Delta\chi^2\simeq 11$ already for modest interaction values.

The physical origin of the rapid deterioration in the SN fit lies in the matter-density balance. The CMB tightly constrains the effective physical matter density before recombination, so transferring energy from CDM to dark energy requires a lower present-day matter abundance $\Omega_{m,0}$. The simultaneous increase in $H_0$ lowers $\Omega_{m,0}$ even further, driving it toward values increasingly disfavored by SN. This mechanism is already present in IDE-only cosmologies but is exacerbated by EDE, which raises $H_0$ before the interaction is introduced and leaves little statistical benefit to be gained from increasing it further. The combined fit therefore gains nothing by moving away from the non-interacting limit.

This mechanism is not specific to the particular coupling or assumptions adopted in our analysis. The same mismatch between the matter abundance constrained by the CMB and the lower present-day value generated by the interaction arises in a broad class of post-recombination models that transfer energy from CDM to dark energy. We therefore expect this limitation to be  general within this class of scenarios. 
We also showed that allowing the dark energy EoS to evolve does not evade this limitation. Although a dynamical EoS can weaken the low-redshift constraints on $\Omega_{m,0}$, perturbative stability typically restricts the branch relevant to increasing $H_0$ to the non-phantom regime. At fixed interaction strength, moving away from $w=-1$ into the non-phantom regime increases the dark energy density in the past, thereby reducing the increase in $H_0$ produced by the interaction.

This restriction on dark sector interactions may also arise when EDE is replaced by other early-time mechanisms that raise $H_0$ by reducing the sound horizon. Provided that the CMB continues to tightly constrain the early-time physical matter density as the interaction is varied, energy transfer from CDM to dark energy will similarly lower the present-day matter abundance, potentially conflicting with SN constraints. Whether this produces a comparable deterioration of the low-redshift fit remains model dependent.

Overall, our results place strong constraints on mixed solutions in which the late-time sector transfers energy from CDM to dark energy. Once an efficient early-time mechanism has absorbed most of the original discrepancy in $H_0$, only a small statistical gain remains for additional late-time dynamics. Such a sector must therefore produce the remaining shift in $H_0$ without degrading the fit to low-redshift observations by more than this limited gain. Given current constraints on the late-time expansion history, satisfying both requirements appears extremely challenging.

\section*{Acknowledgments}
W.G.~and J.S.~acknowledge support from National Aeronautics and Space Administration (NASA) under Grant No.~80NSSC24K0898.~

\appendix

\section{Interacting Dark Energy Equation of State}
\label{app:A}

Our baseline analysis assumes that the EoS of the interacting dark energy fluid remains close to $w=-1$. In this appendix, we test whether allowing the late-time EoS to evolve can qualitatively change our conclusions. We adopt the CPL parametrization, $w(a)=w_0+w_a(1-a)$, and compare five models using the same CMB+DESI+DD+SH0ES dataset combination.\footnote{We adopt flat priors $w_0\in[-3,1]$ and $w_a\in[-3,2]$, with the additional condition $w_0+w_a<0$. For the non-phantom models, we further require $w(a)\geq-1$ for $0\leq a\leq1$. For the non-interacting CPL models, we employ the default
parameterized post-Friedmann (PPF) prescription~\cite{Hu:2007pj,Fang:2008sn}, which permits a regular treatment of the perturbations when $w(a)$ approaches or crosses $-1$.} The best-fit parameters are reported in Table~\ref{tab:cpl_best_fits}. All values of $\Delta\chi^2_{\rm tot}$ are quoted relative to the EDE-only best fit.

\begin{table*}[htpb!]
\centering
\renewcommand{\arraystretch}{1.3}
\resizebox{0.95\textwidth}{!}{%
\begin{tabular}{lccccc}
\toprule
&
\multicolumn{5}{c}{CMB+DESI+DD+SH0ES}
\\
\cmidrule(lr){2-6}
Parameter
& EDE-only
& EDE+IDE
& EDE+IDE-CPL
& EDE+CPL ($w\ge-1$)
& EDE+CPL
\\
\midrule

$100\,\omega_b$
& $2.2599$
& $2.2578$
& $2.2586$
& $2.2586$
& $2.2596$
\\

$\omega_{\rm cdm,0}$
& $0.1274$
& $0.1274$
& $0.1278$
& $0.1278$
& $0.1277$
\\

$100\,\theta_s$
& $1.04128$
& $1.04125$
& $1.04126$
& $1.04126$
& $1.04120$
\\

$\tau_{\rm reio}$
& $0.0614$
& $0.0612$
& $0.0611$
& $0.0611$
& $0.0610$
\\

$\ln(10^{10}A_s)$
& $3.0716$
& $3.0716$
& $3.0722$
& $3.0722$
& $3.0715$
\\

$n_s$
& $0.9825$
& $0.9822$
& $0.9841$
& $0.9841$
& $0.9835$
\\

\hline
$f_{\rm EDE}$
& $0.0904$
& $0.0904$
& $0.0947$
& $0.0947$
& $0.0905$
\\

$\log_{10}z_c$
& $3.526$
& $3.521$
& $3.525$
& $3.525$
& $3.536$
\\

$\theta_i$
& $2.750$
& $2.765$
& $2.777$
& $2.777$
& $2.748$
\\

\hline
$\xi$
& --
& $-3.5\times10^{-4}$
& $-1.0\times10^{-4}$
& $0$
& $0$
\\

$\Delta\Omega_m^{\rm IDE}$
& $0$
& $8.44\times10^{-5}$
& $2.38\times10^{-5}$
& $0$
& $0$
\\

$\Omega_{m,\rm early}^{\rm eff}$
& $0.2996$
& $0.2998$
& $0.3011$
& $0.3010$
& $0.3052$
\\

$\omega_{m,\rm early}^{\rm eff}$
& $0.1507$
& $0.1506$
& $0.1510$
& $0.1510$
& $0.1510$
\\

$w_0$
& --
& $\simeq-1$
& $-0.9884$
& $-0.9884$
& $-0.8387$
\\

$w_a$
& --
& $0$
& $-0.0104$
& $-0.0104$
& $-0.5534$
\\

\hline
$H_0\ [{\rm km\,s^{-1}Mpc^{-1}}]$
& $70.92$
& $70.88$
& $70.82$
& $70.83$
& $70.33$
\\

$r_{\rm drag}\ [{\rm Mpc}]$
& $142.47$
& $142.49$
& $142.26$
& $142.27$
& $142.40$
\\

$\Omega_{m,0}$
& $0.2996$
& $0.2997$
& $0.3011$
& $0.3010$
& $0.3052$
\\

$\sigma_8$
& $0.8328$
& $0.8326$
& $0.8313$
& $0.8311$
& $0.8333$
\\

$S_8$
& $0.8322$
& $0.8322$
& $0.8327$
& $0.8326$
& $0.8405$
\\
\addlinespace

\hline
$\Delta\chi^2_{\rm tot}$
& $0$
& $-0.02$
& $-2.37$
& $-2.41$
& $-8.88$
\\

\bottomrule
\end{tabular}%
}
\caption{Best-fit cosmological parameters for the five models using
CMB+DESI+DD+SH0ES. The EDE+IDE model has a constant fluid equation
of state, $w_0\simeq-1$ and $w_a=0$. The last row reports
$\Delta\chi^2_{\rm tot}\equiv
\chi^2_{\rm tot}(\mathrm{model})-
\chi^2_{\rm tot}(\mathrm{EDE\mbox{-}only})$,
including the common calibration-prior contributions.}
\label{tab:cpl_best_fits}
\end{table*}

The comparison can be summarized as follows:

\begin{itemize}

\item \textbf{EDE-only.}
This model provides the reference, with $H_0=70.92$ km/s/Mpc and $f_{\rm EDE}=0.0904$.

\item \textbf{EDE+IDE.}
This is the baseline interacting model analyzed in the main text. As discussed there, introducing the additional interacting fluid while keeping its EoS close to $w=-1$ does not improve the fit. The best fit lies essentially in the non-interacting limit, with $\xi=-3.5\times10^{-4}$ and $\Delta\Omega_m^{\rm IDE}=8.4\times10^{-5}$, giving only $\Delta\chi^2_{\rm tot}=-0.02$. The preferred value, $H_0=70.88$ km/s/Mpc, is also essentially unchanged relative to EDE-only.

\item \textbf{EDE+IDE-CPL.}
This model extends the baseline EDE+IDE cosmology by allowing the EoS of the interacting fluid to evolve according to the CPL parametrization. As discussed in Sec.~\ref{sec.discussion}, stability of the dark energy perturbations for the branch with energy transfer from CDM to dark energy restricts the fluid to the quintessence-like regime, $w(a)\geq-1$. Allowing both the interaction and the CPL parameters to vary produces only a modest improvement, $\Delta\chi^2_{\rm tot}=-2.37$.\footnote{The EDE+CPL model with $w(a)\geq-1$ is recovered from EDE+IDE-CPL in the limit $\xi\to0$. If this limit were included, the latter model could not have a higher minimum $\chi^2$. However, our prior terminates at $\xi=-10^{-4}$, where the EDE+IDE-CPL best fit is effectively located. The remaining difference, $\Delta\chi^2_{\rm tot}\simeq0.04$, is reproduced by evaluating the EDE+CPL best-fit configuration at this boundary. This confirms that the small difference arises from the finite prior range.} The best fit lies at the prior boundary near the non-interacting limit, with $\xi=-1.0\times10^{-4}$ and $\Delta\Omega_m^{\rm IDE}=2.38\times10^{-5}$, while the EoS itself is driven close to the vacuum-like limit, with $w_0=-0.9884$ and $w_a=-0.0104$.\footnote{For the CPL model we have 
$
\Delta\Omega_m^{\rm IDE}
= -\xi\Omega_{\rm fld}
\int_0^1
a^{-1-3(w_0+w_a)-\xi}
e^{3w_a(a-1)}\,da,
$
which reduces to the constant $w$ expression for $w_a=0$.} The corresponding value, $H_0=70.82\,{\rm km\,s^{-1}\,Mpc^{-1}}$, is slightly lower than in EDE-only. Thus, allowing a dynamical EoS does not generate a significant interacting contribution to the increase in $H_0$.

\item \textbf{EDE+CPL with $w(a)\geq-1$.}
To isolate the effect of the EoS freedom from that of the interaction, we next switch off the interaction and vary the CPL parameters within the same quintessence-like region. The improvement becomes $\Delta\chi^2_{\rm tot}=-2.41$, slightly larger than in EDE+IDE-CPL. The best fit again lies close to $w=-1$, with $w_0=-0.9884$ and $w_a=-0.0104$, and gives $H_0=70.83$ km/s/Mpc. The fact that this non-interacting model performs at least as well as EDE+IDE-CPL shows that the modest gain in the latter is associated with the additional EoS freedom rather than with the interaction itself.

\item \textbf{EDE+CPL with unrestricted $w(a)$.}
Finally, allowing the CPL EoS to explore the full parameter space leads to a substantially larger improvement, $\Delta\chi^2_{\rm tot}=-8.88$. The preferred solution has $w_0=-0.8387$ and $w_a=-0.5534$, corresponding to an EoS that crosses the phantom divide. This improvement, however, is accompanied by a reduction of the Hubble constant to $H_0=70.33$ km/s/Mpc. Thus, the additional late-time freedom can significantly improve the global fit, but the preferred evolution acts in the opposite direction to that required for providing further relief to the Hubble tension.

\end{itemize}

\newpage 

\bibliographystyle{JHEP}

\bibliography{bib}

\end{document}